\documentclass[
 reprint,
 amsmath,amssymb,
 aps,
 floatfix,
 prl,
floatfix,
]{revtex4-1}

\newcommand{\comment}[1]{}

\usepackage[utf8]{inputenc} 
\usepackage{graphicx}
\usepackage{dcolumn}
\usepackage{bm}
\usepackage{subcaption}

\begin{document}


\title{Black Hole Shadows and Invariant Phase Space Structures}

\author{J. Grover}
\email{jai.grover@esa.int}
\author{A. Wittig}
\email{alexander.wittig@esa.int}
\affiliation{
 \small ESA -- Advanced Concepts and Studies Office, European Space Research Technology Centre (ESTEC),Keplerlaan 1, Postbus 299, NL-2200AG Noordwijk, Netherlands}%

\date{\today}

\begin{abstract}
Utilizing concepts from dynamical systems theory, we demonstrate how the existence of light rings, or fixed points, in a spacetime will give rise to families of periodic orbits and invariant manifolds in phase space. It is shown that these structures define the shape of the black hole shadow as well as a number of salient features of the spacetime lensing. We illustrate this through the analysis of lensing by a hairy black hole.
\end{abstract}

\maketitle

\textbf{Introduction.} The study of null geodesic motion can reveal important features of a spacetime. Of particular importance for lensing by compact objects, are the spherical and circular photon orbits (or light rings) that the spacetime admits. The existence of unstable photon orbits around compact objects is associated to multiple images of light sources, and, in the case of black holes, to a shadow in lensing images \cite{Perlick:2004, Bardeen:1973,Falcke:1999pj, Grenzebach:2014}. The orbital frequency and Lyapunov exponents of unstable photon orbits have also been connected to the characteristics of quasi-normal modes for perturbed black holes \cite{Cardoso:2009}.  Stable photon orbits can exist around compact objects, such as boson stars and black holes \cite{Cardoso:2014, Dolan:2016, Cunha:2016}. Their existence has been linked to chaotic scattering in lensing \cite{Cunha:2016, Shipley:2016} and to instabilities of the spacetime \cite{Cardoso:2014}. 

Various methods from dynamical systems have been used previously in the context of general relativity, for example to study chaotic scattering in multi black hole solutions \cite{Cornish:1996de,Dettmann:1994dj,Dettmann:1995ex, Shipley:2016}. Here we focus on studying spacetime lensing in terms of invariant phase space structures such as fixed points, periodic orbits and invariant manifolds. These play a crucial role in dynamical systems where they shape the dynamical behaviour both locally as well as globally \cite{Katok1997}. A prime example in classical mechanics is the circular restricted three body problem (CR3BP) \cite{Szebehely1967}. Extensive analysis, both analytical and numerical, has been done on the Lagrange points, their Lyapunov orbits, and their invariant manifolds \cite{Henon1997,Richardson1980} with major applications in space trajectory design \cite{Gomez2001,Mingotti2009}.

In this work we provide a specific application of this idea, relating the invariant structures from dynamical systems to the fundamental photon orbits of spacetime lensing by (ultra)compact objects. This proves useful when analysing the lensing properties of spacetimes for which the geodesic motion is not completely integrable such as certain hairy black hole solutions \cite{Herdeiro:2014goa}.


\textbf{Lensing setup.} When displaying lensing images we adopt the conventions detailed in \cite{Cunha:2016}, following the framework of \cite{Perlick:2004, Bohn:2014}. This corresponds to a ray-tracing procedure in which the directions an observer can look are related to initial conditions for past-directed null geodesics. These geodesics are then integrated back to a source by integrating the null geodesic equations for the given spacetime. The result is a lensing image showing what the observer sees in any direction on their local sky given some distribution of distant light sources.

The equations of motion of a null geodesic can be derived from the Hamiltonian
$$
H=\frac{1}{2}g^{\mu\nu}(q)p_{\mu}p_{\nu} = 0 \ .
$$
In the following analysis we restrict ourselves to stationary, axi-symmetric spacetimes, with coordinates adapted to the symmetries: $q^{\mu}=(t,r,\theta,\varphi)$, and $p_{\nu} = (p_{t},p_{r},p_{\theta},p_{\varphi})$.

Since $H$ is independent of $t$ and $\varphi$, $p_{t}$ and $p_{\varphi}$ are constants of motion. Let $E=-p_{t}$ and $L=p_{\varphi}$ and introduce the impact parameter $\eta=L/E$.
These constants decouple the motion of $(r,\theta)$ entirely from that of $(t,\varphi)$. The phase space of this system is then the four dimensional $(r,\theta,p_{r},p_{\theta})$.

In these coordinates the Hamiltonian takes the form 
\[
H=g^{rr}p_{r}^{2}+g^{\theta\theta}p_{\theta}^{2}+V(q)\ , 
\]
where we have introduced an effective potential 
\[
V=E^{2}g^{tt}-2ELg^{t\varphi}+L^{2}g^{\varphi\varphi}\ .      
\]


\textbf{Invariant dynamical structures.} A \emph{Hamiltonian dynamical system} is a one parameter flow, in an affine parameter $\lambda$, induced by a vector field $X_H$ as
$$ \dot{x}^a = X_H^{a} = \omega^{ab} \partial_b H \ , $$
where $\omega^{ab}$ is the standard symplectic form, $x^{a} = (q^{\mu}, p_{\nu})$ and $\dot{x}^a=\frac{d}{d\lambda}x^a$.

The most important features of dynamical systems are structures that as a whole remain invariant under the dynamics. These include fixed points, periodic orbits and invariant manifolds. The simplest such structures are the \emph{fixed points} of the dynamics. They greatly influence the behavior of the system in their neighborhood.

In the context of $H$ for axisymmetric spacetimes, the fixed point condition
$\dot{q}=\dot{p}=0$ reduces to
\begin{eqnarray*}
\left.\frac{\partial}{\partial\theta}V\right|_{r_{i},\theta_{i}}  =  0\ , \quad \left.\frac{\partial}{\partial r}V\right|_{r_{i},\theta_{i}}  =  0 \ ,
\end{eqnarray*}
yielding the fixed points $x_i=(r_{i},\theta_{i},0,0)$ in phase space.

These fixed points can be classified by their local stability as 
determined by the eigenvalues $\mu_j$ of the Jacobian $J=\left.D X_H\right|_{r_i,\theta_i}$.
We can distinguish three cases:
\begin{itemize}
\item $\Re(\mu_j)>0$: unstable,
\item $\Re(\mu_j)<0$: stable,
\item $\Re(\mu_j)=0$: center.
\end{itemize}
Due to symplecticity, eigenvalues of $J$ always come in pairs $\mu_j$ and $-\mu_j$.

In purely linear dynamics, the eigenspace corresponding to all stable (unstable) eigenvalues is an invariant linear subspace of points that in forward (backward) time approach the fixed point exponentially. The eigenspace corresponding to the center instead consists of points rotating around the fixed point with angular frequencies related to the imaginary part of the eigenvalue.

This splitting survives in non-linear dynamics. The linear subspaces
generalize to stable, unstable, and center invariant manifolds, respectively, of the same dimension as the corresponding linear subspace \cite[p. 57]{Anosov1988}.
These manifolds are invariant under the dynamics and hence can form separatrices of the dynamics in the system as no trajectories can cross the invariant manifolds. Points in the stable (unstable) manifold exponentially approach the fixed point in forward (backward) time. The dynamics in the center manifold, however, is now determined by the higher order terms of the Hamiltonian.

\textbf{Lyapunov orbits and their invariant manifolds.}
A particular case of center manifold dynamics of a fixed point is given by the
Lyapunov central theorem \cite{Meyer2009}. This states that, under mild non-resonance assumptions, each purely imaginary eigenvalue $\mu_i$ gives rise to a one parameter family $\gamma_\epsilon$ of periodic orbits. For $\epsilon\rightarrow 0$ the orbit $\gamma_\epsilon$ collapses into the fixed point. We refer to this family (or families) of periodic orbits as the Lyapunov family.

\begin{figure}
  \centering
  \includegraphics[width=0.45\textwidth]{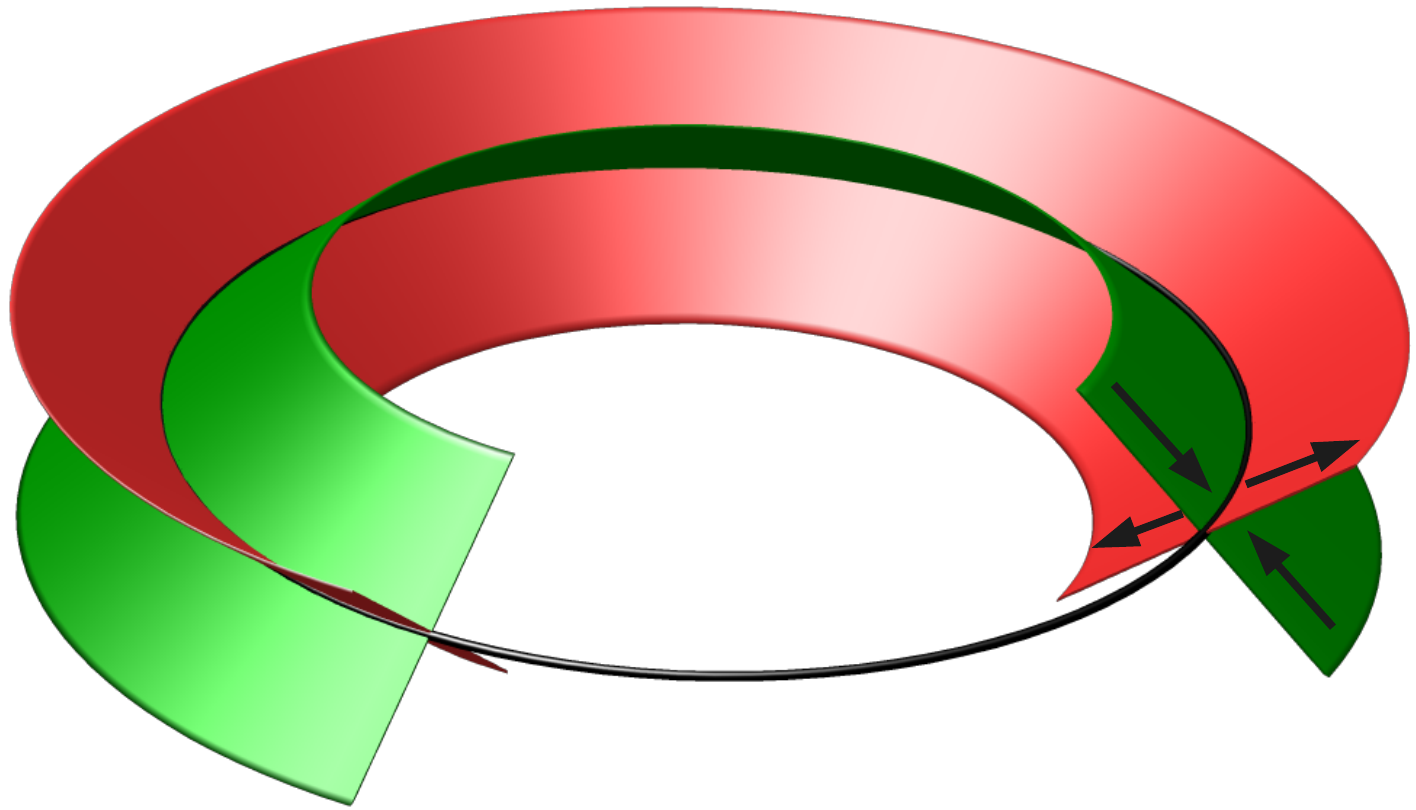}
  \caption{\label{fig:manifolds}Stable (green) and unstable (red) manifolds of a periodic orbit (black).}
\end{figure}

Periodic orbits, whether from a Lyapunov family or not, by themselves
are interesting invariant structures of the dynamics. Similar to the
case of fixed points, they too can have invariant manifolds associated
with them.
Each point on the orbit has an $n$ dimensional manifold attached to it,
which taken over the whole orbit forms an $n+1$ dimensional manifold (see Figure \ref{fig:manifolds}).
As with fixed points, the stable (unstable) invariant manifolds are defined
as those points that in forward (backward) time exponentially spiral in
towards the periodic orbit.

In particular, periodic orbits emanating from a fixed point via a Lyapunov family in some neighborhood around the fixed point exhibit the same manifold structure as the fixed point. A fixed point with a saddle $\times$ center structure thus gives rise to a family of periodic Lyapunov orbits each with a two dimensional stable and unstable manifold attached to it. 

In the context of our problem, those manifolds of the periodic orbits are of great interest due to their higher dimensionality. The fact that $H$ vanishes implicitly defines $p_r$ given $x=(r,\theta,p_{\theta})$ and hence motion can be reduced to three dimensions. This simplifies greatly the visualization and analysis of phase space structures: topologically, the two dimensional manifolds form tubes in the three dimensional reduced phase space, separating it into an inside and an outside.


\textbf{Relation to black hole shadows.} While abstract, these dynamical systems concepts provide a powerful theoretical foundation for understanding the features of strong gravitational lensing by compact objects. To see this it is useful to consider one of the most striking strong lensing features, the emergence of a black hole shadow. This shadow appears as a region of darkness on the lensing image seen by an observer on a black hole background; or, more specifically, to a set of initial conditions for light rays that, when traced backwards from the observer, eventually cross the event horizon of the black hole. 

For certain (stationary) axisymmetric spacetimes, where the motion is completely integrable, implicit analytic expressions defining the shape of the shadow exist \cite{Perlick:2004, Grenzebach:2014}. In particular, it has been shown that the boundary of the black hole shadow corresponds to the set of light rays that inspiral asymptotically onto spherical null orbits around the black hole. The family of such spherical orbits, parameterised by $r$ (or $\eta$), exists over a finite interval of radial positions, and is referred to as a \emph{photon region}. 

A spherical orbit in the photon region oscillates between limiting values of the polar coordinate $\theta$, centered on the equator. The orbits at the extremal radial positions of the photon region are circular - ie they are spherical orbits that never leave the equatorial plane - and are referred to as \emph{light rings}. More generally, a light ring is taken to be any planar circular photon orbit, which implies $\partial_\theta V=\partial_r V=0$. 

This setup suggests an association between the fundamental photon orbits - in particular photon spheres, light rings - (but see also \cite{Cunha:2017, Cornish:1996de, Shipley:2016}) that strongly influence the lensing behaviour, and invariant structures of dynamical systems theory. By definition, light rings are fixed points of the dynamics, while the photon spheres and other periodic orbits will generalise to their Lyapunov orbits. In particular, stable light rings correspond to fixed points with center $\times$ center configuration, while unstable light rings to those of center $\times$ saddle (or saddle $\times$ saddle) type.

It is then natural to see light rings as the fundamental objects from which a range of periodic orbits arise; that the periodic orbits are sometimes spherical is a consequence of the null geodesic equations admitting a fourth constant of motion for the particular classes of solution studied, rather than a general feature.

Making this association has the advantage of providing a formal, and unifying, framework within which to study some features of lensing dynamics, as well as providing a number of useful tools. We will illustrate this through the study of the lensing dynamics of a particular hairy black hole solution.

\begin{figure}[tb]
  \centering
  \includegraphics[width=0.48\textwidth]{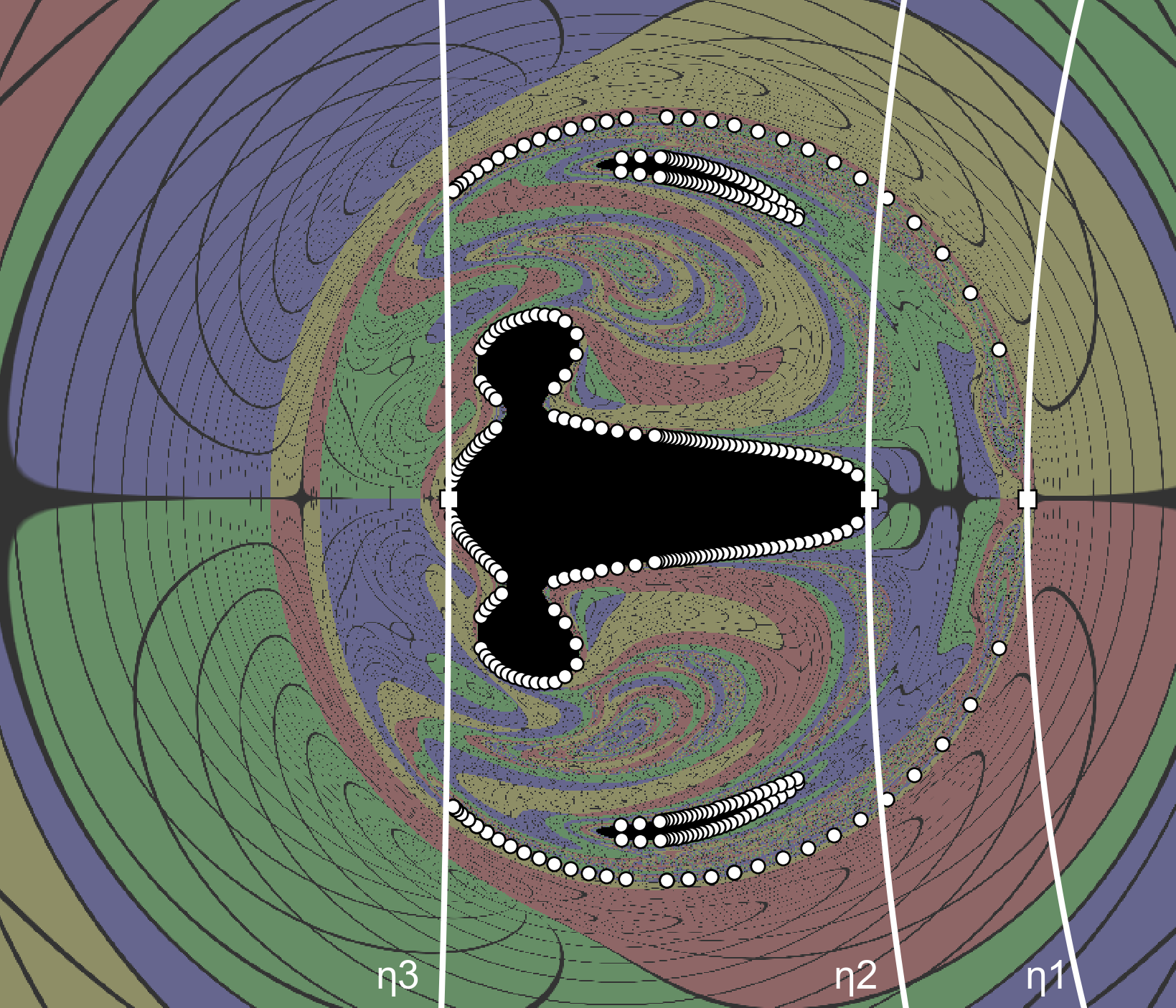}
  \caption{\label{fig:boundary-etas}Intersections of the unstable manifolds of $\mathfrak{L}_1$, $\mathfrak{L}_2$, and $\mathfrak{L}_3$ as well as their Lyapunov orbits with the image plane (white dots). Lines of constant $\eta$ corresponding to $\eta_1$, $\eta_2$, and $\eta_3$ are also shown.}
\end{figure}

\textbf{Case Study: Lensing by a hairy black hole.} For solutions where the motion is not completely integrable the lensing can still be studied numerically, though it then becomes harder to gain insight into the underlying structures for the dynamics. This is the case for a class of stationary, axisymmetric hairy black holes known as Kerr black holes with scalar hair (KBHSH) \cite{Herdeiro:2014goa, Cunha:2015}. We will adopt a particular solution in this class \footnote{Configuration III from \cite{Cunha:2015}} as an example, since it exhibits a number of novel lensing features, such as  multiple disconnected shadows, a principal shadow with non-convex boundary, as well as regions of chaotic motion on the image plane (see Figure \ref{fig:boundary-etas}) \cite{Cunha:2015, Cunha:2016}. As this solution does not have an explicit analytic form, we will apply numerical techniques commonly used in the study of dynamical systems to compute the invariant structures using our raytracing code PyHole \cite{Cunha:2016}.

\begin{table}
\caption{\label{tab:lightrings}Light ring structure of the KBHSH solution.}
\begin{ruledtabular}
\begin{tabular}{cccccc}
   & $\eta$ & X  & Type & Colour & $I^\eta$  \\
\hline
$\mathfrak{L}_1$  & -6.67 & 0.74 & center $\times$ saddle & red & $[-6.67,0.844]$ \\
$\mathfrak{L}_2$ & -4.66 & 0.06 & center $\times$ saddle & green & $[-4.66,-1.87]$ \\
$\mathfrak{L}_3$ & 0.93 & 0.03 & center $\times$ saddle & blue & $[-1.87,0.93]$ \\
$\mathfrak{L}_4$ & 9.72 & 0.30 & center $\times$ center & -- & -- \\
\end{tabular}
\end{ruledtabular}
\end{table}

The light rings of this solution all lie on the equatorial plane and are given in Table \ref{tab:lightrings}. They are labeled $\mathfrak{L}_1,\dots,\mathfrak{L}_4$, with corresponding positions $X_1,\dots,X_4$ \footnote{The compactified radial coordinate $X$ is defined as $X = \frac{X^*}{1+X^*}, \quad X^* = \sqrt{r^2 - r_h^2}$, where $r_h$ is the horizon radius of the black hole.} and $\eta_1, \dots, \eta_4$, and are assigned a colour used to distinguish features arising from different light rings in the following figures.

\begin{figure}
    \centering
    \begin{subfigure}[t]{0.48\textwidth}
        \centering
        \includegraphics[width=0.95\textwidth]{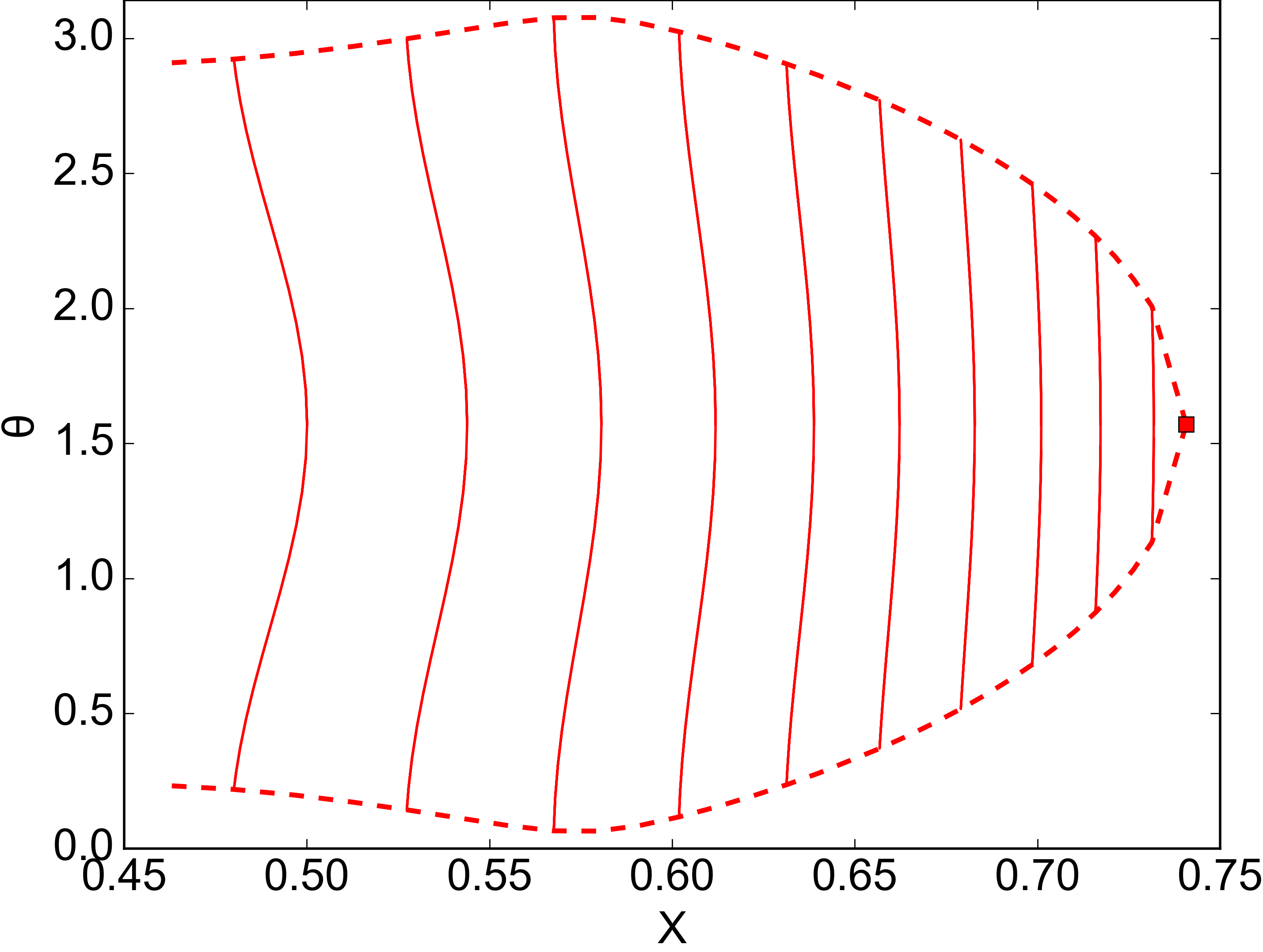}
        \caption{$\mathfrak{L}_1$ (red).}
    \end{subfigure}
    \\[5mm]
    \begin{subfigure}[t]{0.48\textwidth}
        \centering
        \includegraphics[width=0.95\textwidth]{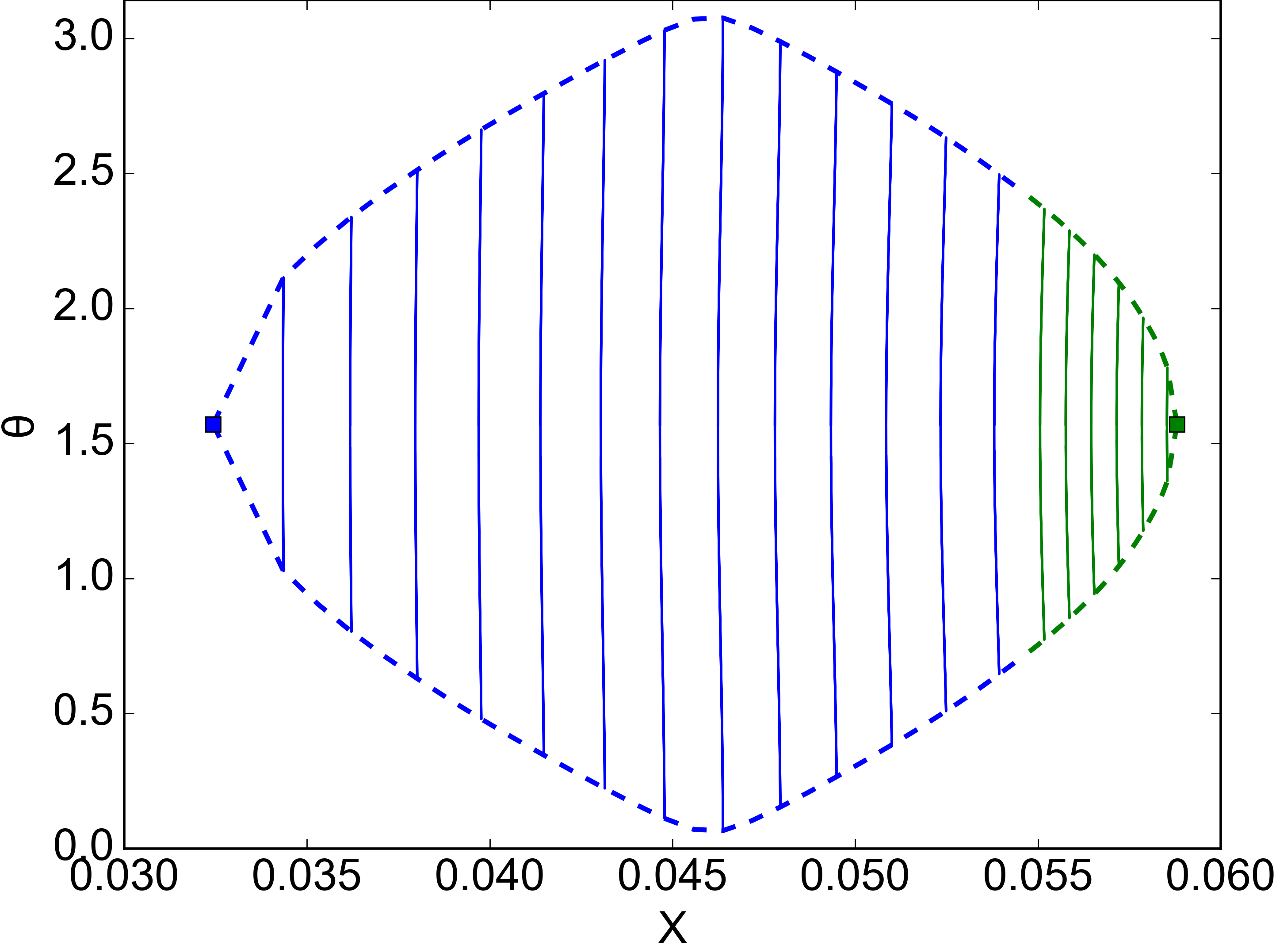}
        \caption{$\mathfrak{L}_2$ (green) and $\mathfrak{L}_3$ (blue).}
    \end{subfigure}
    \caption{\label{fig:boundary-envelope1}Lyapunov families (solid) of light rings (dots) and their envelopes (dashed).}
\end{figure}

The first three light rings $\mathfrak{L}_1, \mathfrak{L}_2$ and $\mathfrak{L}_3$ are of type center $\times$ saddle and thus each gives rise to a family of periodic Lyapunov orbits as shown in Fig. \ref{fig:boundary-envelope1}. Each family can be parametrised by $\eta$ in an interval $I^\eta$. The ranges $I^\eta_1, \dots, I^\eta_3$ for which we identified Lyapunov orbits are also given in Table \ref{tab:lightrings}.

The Lyapunov family associated with $\mathfrak{L}_1$ spans a large range of $\eta$ and orbits clearly deform into non-spherical orbits. Orbits emanating from $\mathfrak{L}_2$ and $\mathfrak{L}_3$, on the other hand, appear nearly spherical. Furthermore, these two families merge into one connected family. This combined family is responsible for generating the black hole shadow.

\begin{figure}[t]
  \centering
  \includegraphics[width=0.48\textwidth]{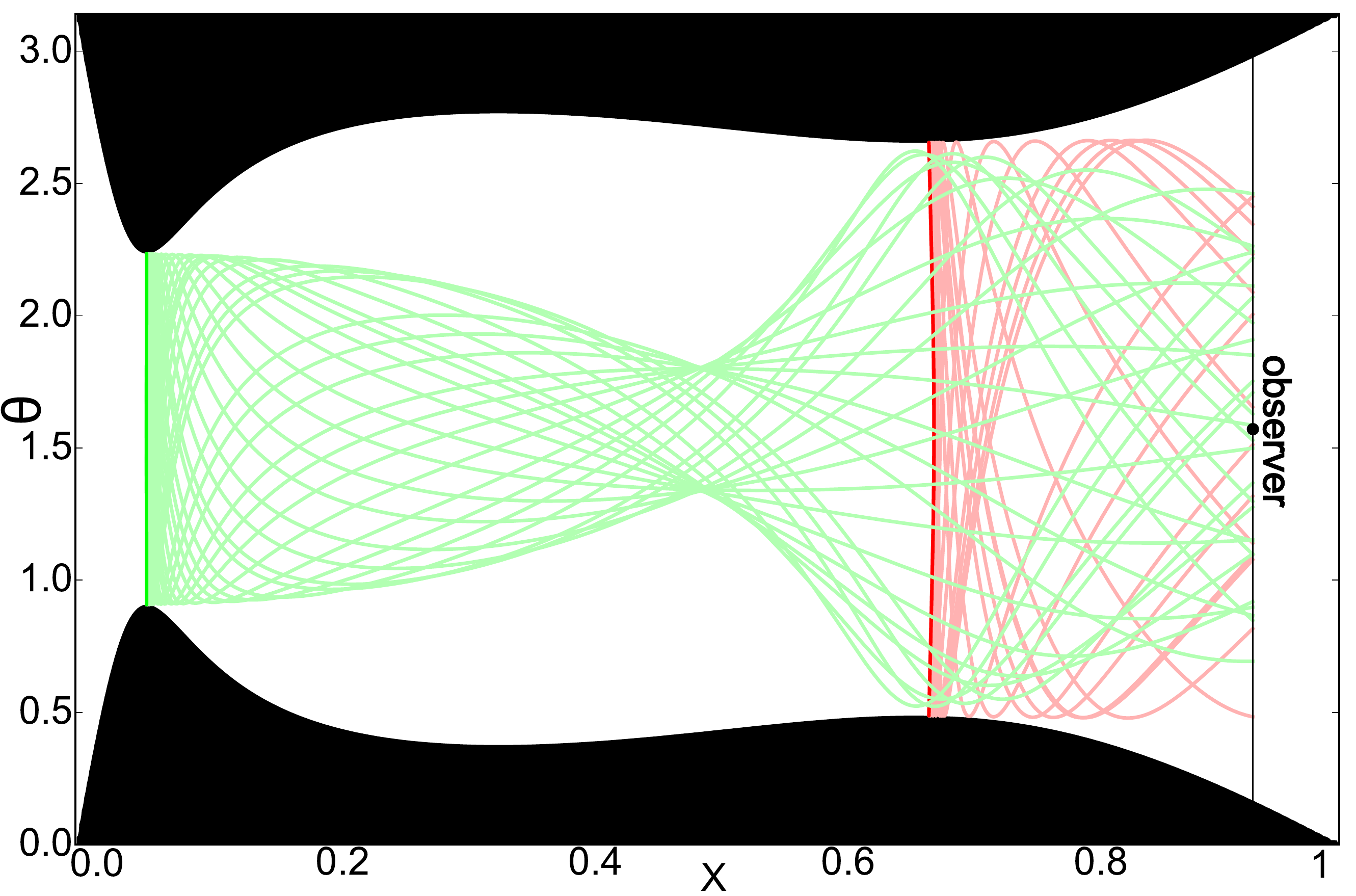}
  \caption{\label{fig:mfd-potential}Projection of the Lyapunov orbits of $\mathfrak{L}_2$ (bright green) and $\mathfrak{L}_1$ (bright red) and their unstable manifolds (light green, light red) for $\eta=-2.6$.}
\end{figure}

\begin{figure}[t!]
    \centering
    \begin{subfigure}[t]{0.48\textwidth}
        \centering
        \includegraphics[width=0.9\textwidth]{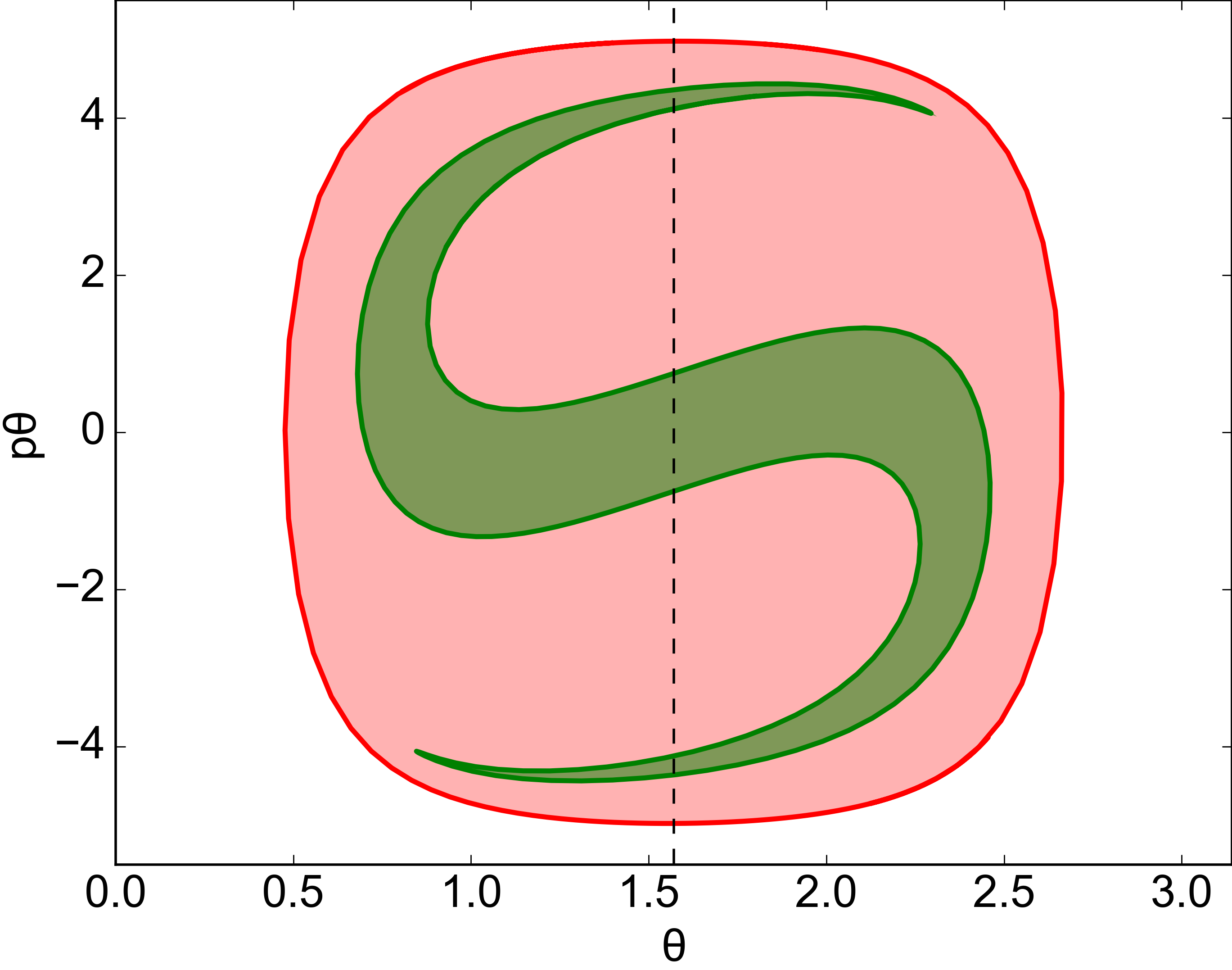}
        \caption{Unstable manifolds of Lyapunov orbits of $\mathfrak{L}_2$ (green) and $\mathfrak{L}_1$ (red) for $\eta=-2.6$.}
    \end{subfigure}
    \\[5mm]
    \begin{subfigure}[t]{0.48\textwidth}
        \centering
        \includegraphics[width=0.9\textwidth]{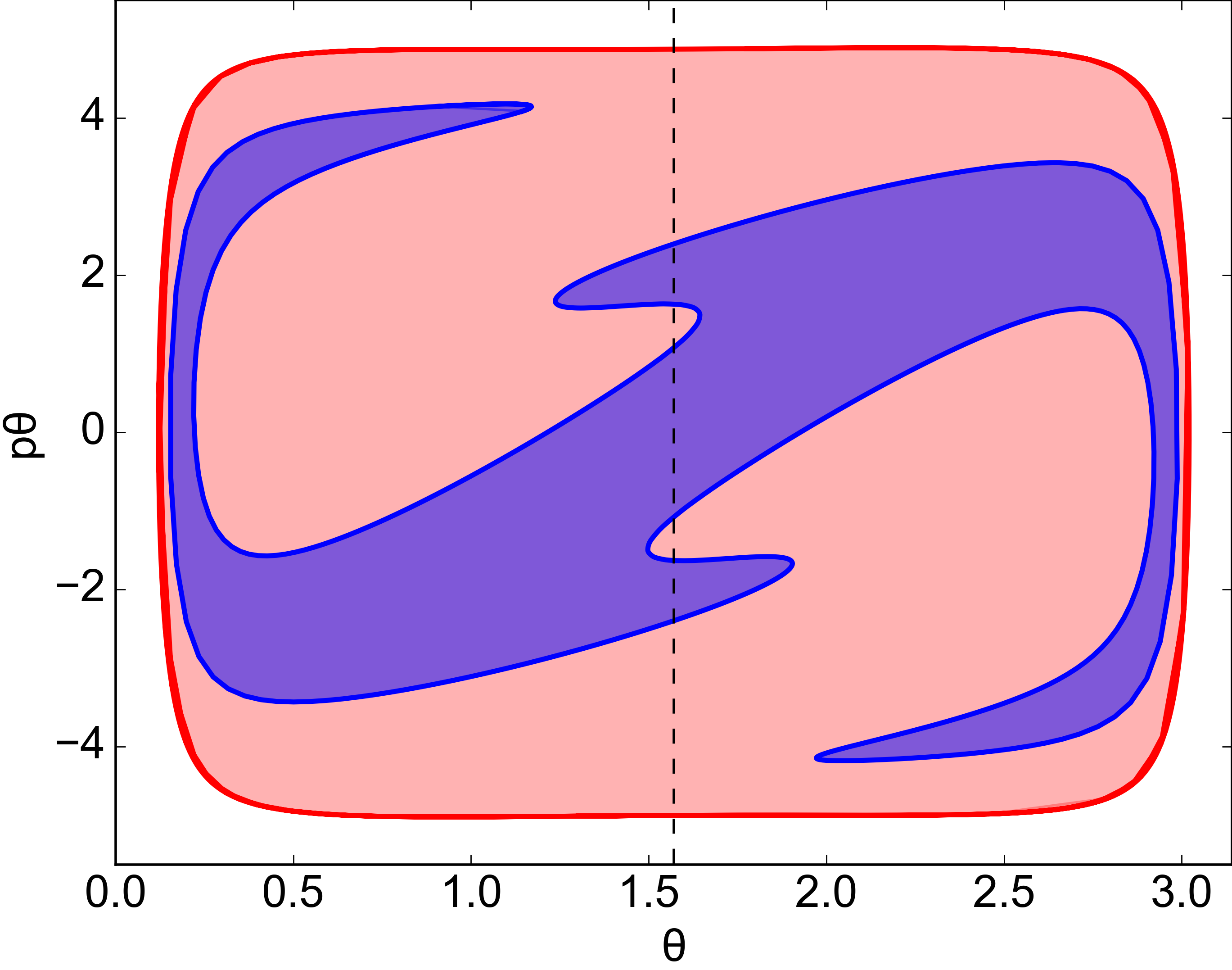}
        \caption{Unstable manifolds of Lyapunov orbits of $\mathfrak{L}_3$ (blue) and $\mathfrak{L}_1$ (red) for $\eta=-0.6$.}
    \end{subfigure}

  \caption{\label{fig:boundary-poincare}Poincaré section in the $r=r_\text{observer}$ plane of unstable manifolds of Lyapunov orbits with different $\eta$.}
\end{figure}

The invariant manifolds of each Lyapunov orbit are two dimensional surfaces  forming tubes in the three dimensional reduced phase space $(r,\theta,p_{\theta})$. This is clearly visible in Figure \ref{fig:mfd-potential}, where we show a projection of orbits in the unstable manifolds of $\mathfrak{L}_1$ and $\mathfrak{L}_2$ along with the effective potential $V$ for $\eta=-2.6$. Both Lyapunov orbits oscillate between the upper and lower boundary of the effective potential. Therefore, to enter the phase space region to their left an orbit must do so from inside the manifold tube. This implies that, in particular, only orbits  inside the unstable manifold tube associated with $\mathfrak{L}_2$ can reach the event horizon for this value of $\eta$.

The shape of these manifolds is further revealed by taking a Poincar\'e section across the manifold tube at the observers radial position. Figure \ref{fig:boundary-poincare} shows the Poincar\'e section in the $(p_{\theta},\theta)$ plane for the same cases of $\eta = -2.6$ and $\eta = -0.6$. For both values there are two Lyapunov orbits originating from $\mathfrak{L}_1$ and either $\mathfrak{L}_2$ or $\mathfrak{L}_3$.

Any trajectory starting within one of the shaded contours is constrained to always move within the corresponding unstable manifold tube. The plots clearly show that the unstable manifolds of Lyapunov orbits of $\mathfrak{L}_2$ and $\mathfrak{L}_3$ lie entirely within the manifold tubes of Lyapunov orbits of $\mathfrak{L}_1$. Furthermore, it is apparent that while the shape of the unstable manifold associated with $\mathfrak{L}_1$ remains simple, the manifolds associated with $\mathfrak{L}_2$ and $\mathfrak{L}_3$ get more and more twisted. This twisting of phase space is caused by stable periodic orbits, emanating from the stable light ring $\mathfrak{L}_4$, which lie outside the manifolds.

To connect these plots with features on the lensing image, note that the intersections of each manifold with the $\theta = \pi/2$ line correspond to trajectories that reach the observer on the equatorial plane. If the observer were to be placed at some other inclination then one would use the corresponding $\theta = \text{const}$ line on the Poincar\'e section to determine which trajectories reach the observer.

Each such intersection point determines the momentum of a light ray that in backward time asymptotically spirals onto the Lyapunov orbit. It also determines a point on the image plane.
Fig. \ref{fig:boundary-etas} shows the lensing image marking the intersection points for a range of values of $\eta$.

\begin{figure}[t!]
  \centering
  \includegraphics[width=0.48\textwidth]{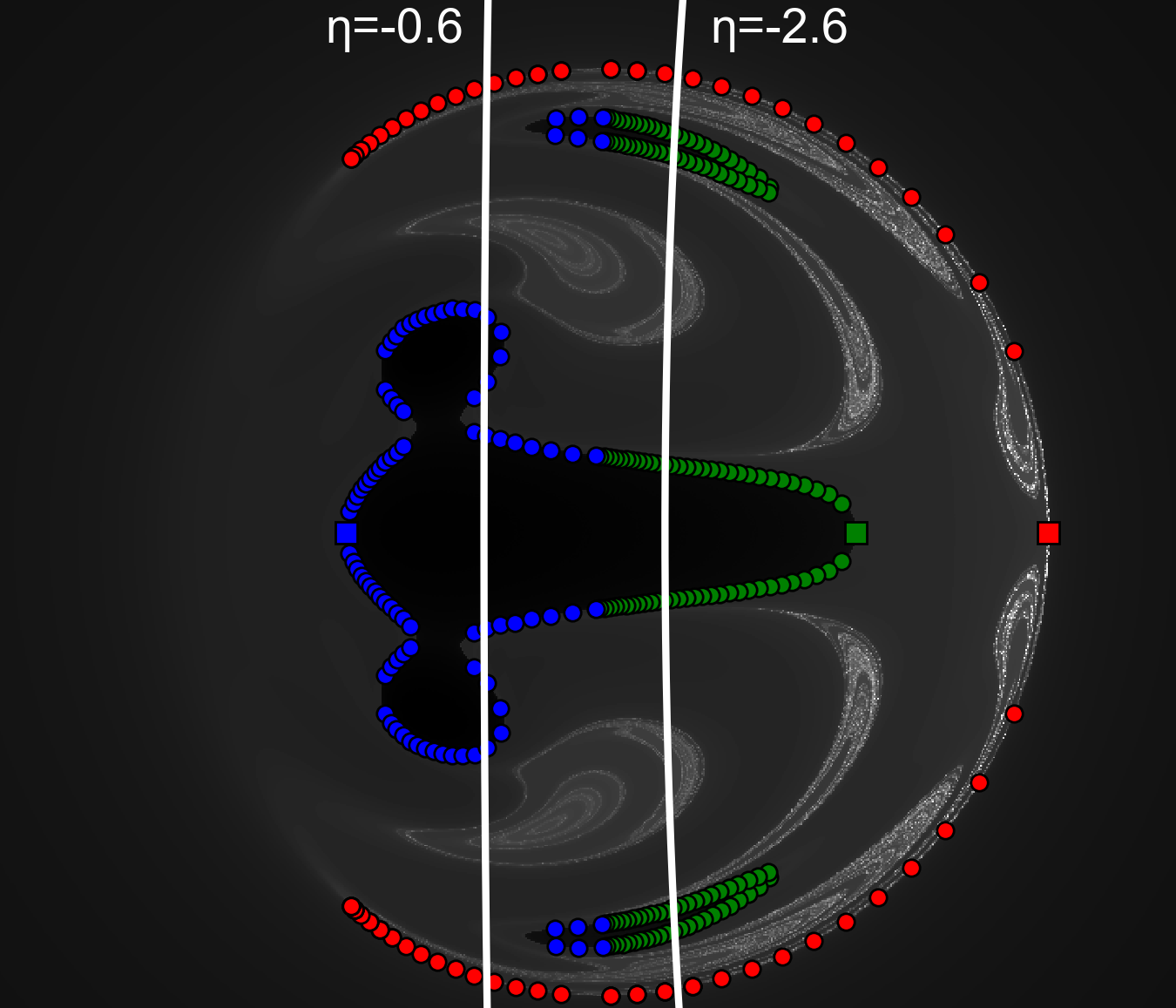}
  \caption{\label{fig:boundary-finaltime-etas}Time delay (logarithmic scale) and unstable manifold intersections with the image plane. Manifolds of $\mathfrak{L}_1$ ($\mathfrak{L}_2, \mathfrak{L}_3$) are marked by red (green, blue) squares; manifolds of their respective Lyapunov orbits by dots of the same color.}
\end{figure}

To better understand this effect, Fig. \ref{fig:boundary-finaltime-etas} shows the the intersection points colored according to their associated light ring. The background is no longer the lensing image, instead each pixel is shaded according to the time delay of each ray (the coordinate time upon arrival at the celestial sphere or event horizon). The time delay is a good heuristic for chaotic motion and also highlights the existence of a 'lensing region' encompassing the chaotic regions and the black hole shadows.

It is apparent from these figures that the boundary of the lensing region is determined by $\mathfrak{L}_1$. This observation is compatible with the analysis in \cite{Cunha:2016} which showed the existence of a pocket in the effective potential that was identified as an important generator of non-trivial dynamics. This pocket only becomes accessible for impact parameters larger than $\eta_1$.

The intersection points deriving from $\mathfrak{L}_2$ and $\mathfrak{L}_3$ instead determine the boundary of the black hole shadow, including its non-convex and disconnected components. For $\eta=-2.6$, for example, the small disconnected shadows are caused by the top and bottom part of the S shape of the Poincar\'e section (Figure \ref{fig:boundary-poincare}), while the main shadow corresponds to the central part of the shape. Furthermore, it is evident that all points inside the manifold tube correspond to points inside the shadow.


\textbf{Conclusion.} Applying techniques from dynamical systems theory to spacetime lensing yields novel insights into the underlying phase space structures governing the motion along null geodesics. In particular we have shown that the invariant manifolds of certain Lyapunov orbits are directly related to black hole shadows even in the case of complicated non-convex, disconnected shadows. All of these structures arise naturally from the existence of fixed points (light rings).

This formalism also provides us with an alternative procedure for computing the boundaries of black hole shadows. It can be implemented using standard techniques from numerical dynamical systems. While the shadow analysis is naturally restricted to black hole solutions, the same dynamical systems framework is also applicable to boson stars and other horizonless solutions, since they too can exhibit stable and unstable light rings, leading to interesting lensing dynamics.

\bibliography{references}

\end{document}